\documentclass[12pt]{article}  
\begin{document}

\newcommand{\sheptitle}
{New uncertainties in QCD-QED rescaling factors using  Quadrature Method }
\newcommand{\shepauthor}
{Mahadev Patgiri${^\dag}$\footnote{corresponding author: {\it{E-mail:}} mahadev@scientist.com}   
and N.Nimai Singh${^\ddag}$\footnote{ICTP Regular Associate; E-mail:nimai03@yahoo.com} }

\newcommand{\shepaddress}
{${^\dag}$Department of Physics, Cotton College, Guwahati-781001, India\\
${^\ddag}$ Department of Physics, Gauhati University, Guwahati-781014, India, and The Abdus Salam ICTP, Trieste,Italy.}

\newcommand{\shepabstract}
{In this paper we briefly outline the quadrature method for estimating uncertainties in
 a function which depends on several variables,  and  apply  it to estimate  the numerical 
uncertainties in QCD-QED rescaling factors. We employ here the one-loop order in QED and three-loop
 order in QCD evolution equations of the fermion mass renormalization. 
Our present calculation is found to be new and 
also reliable   compared to the earlier values employed by various authors.}

\begin{titlepage}
\begin{flushright}
hep-ph/0406044
\end{flushright}
\begin{center}
{\large{\bf\sheptitle}}
\bigskip\\
\shepauthor
\\
\mbox{}\\
{\it\shepaddress}\\
\vspace{.5in}
{\bf Abstract}
\bigskip
\end{center}
\setcounter{page}{0}
\shepabstract
\end{titlepage}
\section{Introduction}
 An important aspect of particle physics is the concept of mass of a fermion. The experimental values of the fermion masses  
are the  input values that we always take in various calculations, and they  are generally known as physical masses of the fermions. 
Mathematically such  physical mass  is  the running mass $m_{f}(\mu)$ of the fermion defined at the scale equal to its own  
physical mass, i.e., $m_{f}(\mu=m_{f})=m_{f}(m_{f})$. This is true for heavier quarks $f=t,b,c$  
which have  $m_{f}(m_{f})>1$GeV  and for charged leptons $(\tau,\mu, e)$. However for lighter quarks $(f=s,d,u)$ which have 
 masses  lesser than $1$GeV,
the physical mass is defined as $m_{f}(1GeV)$. On the other hand, the fermion masses predicted in Grand unified theories
 (GUTs)[1], are defined at the grand unification scale ${M_{X}}$. In order to express these masses at  lower  energy scales
we need  the   renormalisation of fermion masses from high scale to low scale[2].
At  low energy scale below the top-quark mass scale ${m_{t}}$, the fermion mass renormalisation is
governed by the QCD-QED symmetry gauge group ${SU(3)_{C}\times U(1)_{em}}$. Thus
the fermion mass renormalisation from top-quark mass scale down to the physical quark mass scale
is parametrised by the QCD-QED rescaling factors ${\eta_{f}}$ defined by 
$\eta_{f}=m_{f}(m_{f})/m_{f}(m_{t})$. This is an important parameter which appears in many expressions
related to the predictions of GUTs at low energy scale. The calcualtion of  ${\eta_{f}}$
involves the input values  of e.m and strong gauge coupling constants  ${\alpha}$ and ${\alpha_{3}}$.
The uncertainties associated with the experimental values of ${\alpha}$ and ${\alpha_{3}}$ will  propagate through
 several intermediate scales upto low energies in the definition of QCD-QED rescaling factors, 
making the uncertainties larger and larger.
In this respect, a careful estimation of uncertainties
in ${\eta_{f}}$ using a reliable method which can control the magnifying 
tendency of the uncertainties in ${\eta_{f}}$, is  highly desirable. 
The  uncertainties in ${\eta_{f}}$ so far reported  in the literature[4,12], do 
not match with each other,   and the methods adopted by them  are also not clearly specified.
In this context we find the quadrature method[3] for the  estimation
of uncertainties of a function which depends on several variables, to be quite satisfactory and appropriate.
 It will be an important numerical exercise  to estimate the uncertainties in $\eta_{f}$ using the quadrature method 
and compare the results with the earlier values. We shall employ the one-loop order in QED and three-loop order in QCD 
evolution equations of the fermion mass renormalization.

The paper is organised as follows. In Section 2, we outline the procedure for the quadrature method. 
Section 3 is devoted to a brief note on how to define QCD-QED rescaling factors and related quantities[4,5,6]. In Section 4,
we present the numerical calculations of rescaling factors and their respective uncertainties.  The paper concludes with a 
summary in Section 5. 

\section{\bf Estimation of uncertainties in Quadrature Method }

We define the quadrature method in the following way.
If a function F depends on several variables ${x_{i}}$, where i=1,2,...n, then the uncertainties
in F resulting from the uncertainties of the independent variables ${x_{i}}$ can be estimated by the following expression,
\begin{equation}
\delta F= \pm \sqrt{\sum\left(\frac{\partial F}{\partial x_{i}}\right)^{2}(\delta x_{i})^{2}}
\end{equation}
where ${\pm \left(\frac{\partial F}{\partial x_{i}}\right)(\delta x_{i})}$ is the error in F due to the error in ${i^{th}}$ variable
${x_{i}}$. Here the uncertainties (or random errors of independent variables)  add in quadrature[3]  like orthogonal vectors, 
 ${\vec{a}}$ and ${\vec{b}}$ whose resultant magnitude is given by 
${\sqrt{a^{2}+b^{2}}}$.

We give here three simple properties  which will be needed  to  estimate  the uncertainties in different quantities leading 
to the calculations of rescaling factors and also other parameters which depend on it (e.g.,neutrino masses and mixings):\\
\\
{\bf (i)}: If ${F(x,y)=f(x,y)\times g(x,y)}$, the uncertainties in F due to the random errors ${\pm\delta x}$ and  ${\pm\delta y}$
in the independent variables x and y respectively, are given by
\begin{equation}
\delta F = \pm \sqrt{\left(g\frac{\partial f}{\partial x}+f\frac{\partial g}{\partial x}\right)^{2}(\delta x)^{2}
+\left(g\frac{\partial f}{\partial y}+f\frac{\partial g}{\partial y}\right)^{2}(\delta y)^{2}}
\end{equation}

{\bf (ii)}: If ${F(x,y)=\frac{f(x,y)}{g(x,y)}}$, the uncertainties in F due to the random errors ${\pm\delta x}$ and  ${\pm\delta y}$
in the independent variables x and y respectively, are given by
\begin{equation}
\delta F = \pm \sqrt{\left(\frac{1}{g}\frac{\partial f}{\partial x}-\frac{f}{g^{2}}\frac{\partial g}{\partial x}\right)^{2}(\delta x)^{2}+\left(\frac{1}{g}\frac{\partial f}{\partial y}-\frac{f}{g^{2}}\frac{\partial g}{\partial y}\right)^{2}(\delta y)^{2}}
\end{equation}

{\bf (iii)}: If ${F(x,y)=\left(\frac{f(x)}{g(y)}\right)^{p}}$, then the uncertainties in F due to the uncertainties
${\pm\delta x}$ and  ${\pm\delta y}$ in the independent variables x and y respectively, are
\begin{equation}
\delta F = \pm F p\sqrt{\frac{1}{f^{2}}\left(\frac{\partial f}{\partial x}\right)^{2}(\delta x)^{2}+\frac{1}{g^{2}}\left(\frac{\partial g}{\partial y}\right)^{2}(\delta y)^{2}}
\end{equation}

\section{\bf QCD-QED rescaling factors}

The QCD-QED rescaling factor ${\eta_{f}}$ of the fermion f , which can take care of
the fermion-mass-renormalisation from the top-quark mass scale down to the physical fermion mass scale
 ${m_{f}(m_{f})}$, is defined as [7] 
\begin{equation}
\eta_{f}= \frac{m_{f}(m_{f})}{m_{f}(m_{t})}
=\left [\frac{m_{f}(m_{f})}{m_{f}(m_{t})}\right ]_{QED}\times \left [\frac{m_{f}(m_{f})}{m_{f}(m_{t})}\right ]_{QCD}
\end{equation}
for f = b, c quarks having  ${m_{f}(m_{f})}>1$ GeV,

\begin{equation}
\eta_{f}= \frac{m_{f}(1GeV)}{m_{f}(m_{t})}
=\left [\frac{m_{f}(1GeV)}{m_{f}(m_{t})}\right ]_{QED}\times \left [\frac{m_{f}(1GeV)}{m_{f}(m_{t})}\right ]_{QCD}
\end{equation}
for f = s, d, u quarks having ${m_{f}(m_{f})}<1$ GeV, and

\begin{equation}
\eta_{f}= \left [\frac{m_{f}(m_{f})}{m_{f}(m_{t})}\right ]_{QED}, f= \tau, \mu, e
\end{equation}
Here ${m_{f}(m_{t})}$ is the running mass of the fermion f at the top-quark mass scale,
${\mu=m_{t}}$, and ${m_{f}(m_{f})}$ corresponds to the physical mass of the fermion.
For convenience we also define a quantity ${R^{f}(\mu, \mu ^{\prime})}$ which represents the inverse
of the QCD-QED rescaling factor in the narrow range, ${\mu ^{\prime}-\mu}$, ${(\mu > \mu ^{\prime})}$,

$${ R^{f}(\mu, \mu ^{\prime})
=\left [\frac{m_{f}(\mu)}{m_{f}(\mu ^{\prime})}\right ]_{QED}\times \left 
[\frac{m_{f}(\mu)}{m_{f}(\mu ^{\prime})}\right ]_{QCD}}$$
\begin{equation}
= R^{f}_{QED}(\mu, \mu ^{\prime}) \times R^{f}_{QCD}(\mu, \mu ^{\prime})
\end{equation}
The definitions of ${\eta_{f}}$ and ${R^{f}}$ in Eqs. (5)-(8), require only the contributions from the 
QED part for the charged leptons. Making use of Eq.(8) for successive narrow mass ranges: 
${m_{b}}$-${m_{t}}$, ${m_{\tau}}$-${m_{b}}$,
${m_{c}}$-${m_{\tau}}$, 1GeV-${m_{c}}$, ${m_{s}}$-1GeV, ${m_{\mu}}$-${m_{s}}$, ${m_{d}}$-${m_{\mu}}$,
${m_{u}}$-${m_{d}}$, ${m_{e}}$-${m_{u}}$, the QCD-QED rescaling factor  ${\eta_{f}}$ defined in Eqs.(5)-(7), can
be rewritten as
$${\eta_{b} = \eta_{b}(m_{b}, m_{t}) = 1/R^{b}(m_{b}, m_{t}),}$$
$${\eta_{\tau} = \eta_{\tau}(m_{\tau}, m_{t}) = 1/\left [R^{l}_{QED}(m_{\tau}, m_{b})R^{l}_{QED}(m_{b}, m_{t})\right ],
 l= \tau, \mu, e,}$$
$${\eta_{c} = \eta_{c}(m_{c}, m_{t}) = 1/\left [R^{c}(m_{c}, m_{\tau})R^{c}(m_{\tau}, m_{b})R^{c}(m_{b}, m_{t})\right ],}$$
$${\eta_{s,d} = \eta_{s,d}(1GeV, m_{t}) = \eta_{b}/\left [R^{s,d}(1GeV, m_{c})R^{s,d}(m_{c}, m_{\tau})R^{s,d}(m_{\tau}, m_{b})
\right ],}$$ 
$${\eta_{u} = \eta_{u}(1GeV, m_{t}) = \eta_{c}/\left [R^{u}(1GeV, m_{c})\right ],}$$
$${\eta_{\mu} = \eta_{\mu}(m_{\mu}, m_{t}) = \eta_{\tau}/\left [R^{l}_{QED}(m_{\mu}, m_{s})R^{l}_{QED}(m_{s}, 1GeV)\right ],}$$
\begin{equation}
\eta_{e} = \eta_{e}(m_{e}, m_{t}) = \eta_{\mu}/\left [R^{l}_{QED}(m_{e}, 
m_{u})R^{l}_{QED}(m_{u}, m_{d})R^{l}_{QED}(m_{d},m_{\mu})\right ]\end{equation}
We use one-loop order in QED and three-loop order in QCD evolution equations of fermion mass renormalistion
for the evaluations of ${R^{f}_{QED}}$ and  ${R^{f}_{QCD}}$ respectively. The contribution of one-loop QED, 
running from the scale ${\mu^{\prime}}$ to ${\mu}$, to the rescaling factors through Eq.(8) is now given by[7]
\begin{equation}
{ R^{f}_{QED}(\mu^{\prime}, \mu})=\left [ \frac{\alpha(\mu)}{\alpha(\mu^{\prime}}\right ]^{r_{f}^{QED}},
\mu > \mu^{\prime},
\end{equation}
where
$${r_{f}^{QED}(\mu^{\prime},\mu) = \gamma_{0}^{QED}/b_{0}^{QED},}$$
$${\gamma_{0}^{QED} =-3Q_{f}^{2},}$$
\begin{equation}
b_{0}^{QED} = \frac{4}{3}\left[3\sum Q_{u}^{2}+3\sum Q_{d}^{2}+ \sum Q_{e}^{2}\right].
\end{equation}
The summation in Eqs.(11) is over the active fermions at the relevant mass scale, and f is the specific
fermion under consideration. We employ the one loop RGE for the estimation of e.m. gauge couplings
${\alpha(\mu)}$ at successive renormalisation points,
\begin{equation}
\frac{1}{\alpha(\mu^{\prime})} = \frac{1}{\alpha(\mu)} + \frac{b_{0}^{QED}}{2\pi}ln\left(\frac{\mu}{\mu^{\prime}}\right)
\end{equation}
The three-loop QCD running quark mass formula is given by [7,8]
\begin{equation}
m_{q}(\mu) = \hat{m_{q}}\left[b_{0}\frac{\alpha_{3}(\mu)}{2\pi}\right]^{(\gamma_{0}/b_{0})} 
\times \left( 1+A\frac{\alpha _{3}(\mu)}{4\pi}+B\left[\frac{\alpha_{3}(\mu)}{4\pi}\right]^{2}\right)
\end{equation}
where ${\hat{m_{q}}}$ is the scale-invariant quark mass, and
$${A = \frac{\gamma_{1}}{b_{0}}- \frac{b_{1}\gamma_{0}}{b_{0}^{2}},}$$
\begin{equation}
B= \frac{1}{2} \left[A^{2}+ \frac{\gamma_{2}}{b_{0}}+ \frac{b_{1}^{2}\gamma_{0}}{b_{0}^{3}}
- \frac{b_{1}\gamma_{1}}{b_{0}^{2}} - \frac{b_{2}\gamma_{0}}{b_{0}^{2}}\right]
\end{equation}

The QCD ${\beta}$ -functions and anomalous dimensions are given as,
$${\gamma_{0} = 4,}$$ 

$${\gamma_{1} = \frac{202}{3}-\frac{20}{9}n_{f},}$$

$${\gamma_{2} = \frac{3747}{3}-\left(\frac{160}{3}\xi (3)+\frac{2216}{27}\right)n_{f}-\frac{140}{81}n_{f}^{2},}$$ 

$${b_{0} = 11-\frac{2}{3}n_{f},}$$

$${b_{1} = 102-\frac{38}{3}n_{f},}$$

\begin{equation}
b_{2} = \frac{2857}{2}-\frac{5033}{18}n_{f}+\frac{325}{54}n_{f}^{2}, 
\end{equation}
where ${n_{f}}$ is the number of quark flavours at the relevant mass scale and 
${\xi(3) =1.202}$. The QCD rescaling contribution to ${R^{f}}$ in the relevant mass-range
(${\mu^{\prime}}$-${\mu}$) defined in Eq.(8), can be obtained from Eq.(13) as

\begin{equation}
  R^{f}_{QCD}(\mu, \mu ^{\prime}) = \left[\frac{\alpha_{3}(\mu)}{\alpha_{3}(\mu^{\prime})}\right]^{(\gamma_{0}/b_{0})} 
\times \frac{1+A\frac{\alpha _{3}(\mu)}{4\pi}+B\left[\frac{\alpha_{3}(\mu)}{4\pi}\right]^{2}}
{1+A\frac{\alpha _{3}(\mu^{\prime})}{4\pi}+B\left[\frac{\alpha_{3}(\mu^{\prime})}{4\pi}\right]^{2}}
\end{equation}
since ${\alpha_{3}(\mu)}$ changes smoothly with ${\mu}$, so values of the constants A and B remain
as same for a given energy range including the limits.

Using Eqs.(10) and (14) via (8), the QCD-QED rescaling factors can be estimated.The values of 
${\alpha_{3}(\mu)}$ in Eq.(16) can be obtained by solving the three-loop QCD RGE with ${(\alpha_{3}= g_{3}^{2}/4\pi)}$,[9]
\begin{equation}
\mu\frac{dg_{3}(\mu)}{d\mu} = \beta(g_{3}(\mu))=-\frac{b_{0}}{16\pi^{2}}g_{3}^{3}+\frac{b_{1}}{(16\pi^{2})^{2}}g_{3}^{5}
-\frac{b_{2}}{(16\pi^{2})^{3}}g_{3}^{7}
\end{equation}
The conventional solution of Eq.(17) having a constant of integration called QCD dimensional parameter 
${\Lambda}$-which provides a parametrisation of the ${\mu}$ dependence of ${\alpha_{3}(\mu)}$ [9]is given by

$${\alpha_{3}(\mu)=\frac{4\pi}{\beta_{0}ln(\mu^{2}/\Lambda^{2})}[1-\frac{2\beta_{1}}{\beta_{0}^{2}}
\frac{ln[ln(\mu^{2}/\Lambda^{2})]}{ln(\mu^{2}/\Lambda^{2})}+\frac{4\beta_{1}^{2}}{\beta_{0}^{4}ln^{2}(\mu^{2}/\Lambda^{2})}\times}$$
\begin{equation}
\left(\left(ln[ln(\mu^{2}/\Lambda^{2})]-\frac{1}{2}\right)^{2}+\frac{\beta_{2}\beta_{0}}{8\beta_{1}^{2}}-\frac{5}{4}\right)]
\end{equation}
where
\begin{equation}
\beta_{0}=b_{0}, \beta_{1}=b_{1}/2, \beta_{2}=2b_{2}.
\end{equation}
Since ${\beta}$-function coefficients change by discrete amount as flavour thresholds are crossed while integrating
the differential equation(17) for ${\alpha_{3}(\mu)}$, ${\Lambda}$ also changes to take care of the validity of
Eq.(18) for all values of ${\mu}$. This leads to the concept of different ${\Lambda^{(n_{f})}}$ for each range of ${\mu}$
corresponding to an effective number of quark flavours ${n_{f}}$. In the ${\bar{MS}}$ scheme, one finds the relation among different
 ${\Lambda^{(n_{f})}}$ [10] as

$${\Lambda^{(4)}_{MS}\simeq \Lambda^{(5)}_{MS}\left[m_{b}/\Lambda^{(5)}_{MS}\right]^{2/25}\left[2ln\left[m_{b}/\Lambda^{(5)}_{MS}\right]\right]^{(963/14375)}}$$
\begin{equation}
\Lambda^{(4)}_{MS}\simeq \Lambda^{(3)}_{MS}\left[\Lambda^{(3)}_{MS}/m_{c}\right]^{2/25}
\left[2ln\left[m_{c}/\Lambda^{(3)}_{MS}\right]\right]^{-(107/1875)}
\end{equation}
The QCD parameter ${\Lambda^{(5)}_{MS}}$ which corresponds to ${n_{f}=5}$ can be calculated by using Eq.(18) with the input
experimental value of ${\alpha_{3}(M_{Z})}$ and then  ${\Lambda^{(4)}_{MS}}$ and  ${\Lambda^{(3)}_{MS}}$ can be obtained from
Eq.(20) through a search programme. Subsequently, ${\alpha_{3}(\mu)}$ and ${R^{f}_{QCD}}$ can be calculated for the energy,
${\mu < M_{Z}}$.\\

If we consider SUSY above ${M_{Z}}$, 
assuming the existence of one-light Higgs doublet (${N_{H}}$) and five quark flavours (${n_{f}=5}$)
in the energy range ${m_{t}-M_{Z}}$, the strong gauge coupling at the scale ${m_{t}}$ are evaluated using RGE solution with 
one loop order
\begin{equation}
\frac{1}{\alpha_{3}(\mu)} = \frac{1}{\alpha_{3}(M_{Z})} + \frac{3}{2\pi} ln\frac{\mu}{M_{Z}} 
\end{equation}
for ${\mu > M_{Z}}$.

\section{\bf {Numerical calculations and results}:}
For numerical calculation we take the input values of fermion masses as [9]\\
$${m_{t}=175 GeV, m_{b}=4.25 GeV, m_{\tau}=1.785 GeV,}$$
$${m_{c}=1.25 GeV, m_{s}=0.18 GeV, m_{\mu}=0.105 GeV,}$$
\begin{equation}
{m_{d}=0.009 GeV}, {m_{u}=0.005 GeV}, {m_{e}=.00051 GeV},
\end{equation}
and CERN-LEP data [11] of ${\alpha^{-1}(M_{Z})=127.9\pm0.1}$ and strong coupling constant
${\alpha_{3}(M_{Z})=0.1172\pm .002}$ at ${M_{Z}=91.187 GeV}$, which is referred to as Case-I.
For comparison, we also estimate the uncertainties using values of ${\alpha_{3}(M_{Z})=0.120\pm .0028}$ referred to as Case-II
and ${\alpha_{3}(M_{Z})=.118\pm .007}$ made use for earlier calculation of uncertainties [12], now referred to as Case-III.

 Now using above data, we evaluate the gauge couplings ${\alpha(\mu)}$  and ${\alpha_{3}(\mu)}$ at various renormalisation
points starting from ${m_{t}}$ down to the individual fermion mass (for quark it stops at 1GeV) 
from their corresponding Eqs.(12),(18)and (21) and values are presented in Table-1 and Table-5 respectively. 
The co-efficients of ${\beta}$-functions and anomalous dimensions in RGEs for QED and QCD with constants A, B
 relevant in different energy ranges are estimated and shown in Tables-1,2,3. We also calculate the 
inverse of the fermion mass renormalisation factors ${R_{f}^{QED}(\mu, \mu^{\prime})}$ 
for charged leptons, up-quarks and down-quarks shown in Table-1,2 and ${R_{f}^{QCD}(\mu, \mu^{\prime})}$ in three Cases-I,
II, III in Table-5. The values of QCD dimensional parameters ${\Lambda^{f}_{MS}}$ for estimating ${\alpha_{3}(\mu)}$ at various
energy scales are presented in Table-4.

 We present below the numerical values of QCD-QED rescaling factors ${\eta_{f}}$ (f=b, c, s, d,u) with uncertainties
for different values of ${\alpha_{3}(M_{Z})}$. It is observed that the uncertainties in the input values of ${\alpha_{3}(M_{Z})}$
propagate down to the lower energy scale, causing the uncertainties in ${\eta_{f}}$ larger and larger:\\

\begin{center}
\begin{tabular}{cccccc}\hline
Case & ${\alpha_{3}(M_{Z})}$ & ${\eta_{b}}$ & ${\eta_{c}}$ & ${\eta_{s,d}}$ & ${\eta_{u}}$\\
\hline
\\
I & ${0.1172\pm .002}$ &  ${1.5380_{-0.0335}^{+0.0361}}$  & ${2.1785_{-0.1580}^{+0.2186}}$
&  ${2.4329_{-0.2182}^{+0.2607}}$ &  ${2.4489_{-0.2186}^{+0.2607}}$\\
\\
\hline
\\
     
II & ${0.1200\pm .0028}$ &  ${1.5649_{-0.0479}^{+0.0503}}$  & ${2.2487_{-0.2067}^{+0.2337}}$
&  ${2.6486_{-0.3620}^{+0.4491}}$ &  ${2.6660_{-0.3844}^{+0.4239}}$\\
\\
\hline
\\
III & ${0.1180\pm .007}$ &  ${1.5458_{-0.1165}^{+0.1282}}$  & ${2.2110_{-0.4548}^{+0.6298}}$
&  ${2.5106_{-0.7162}^{+1.1991}}$ &  ${2.5271_{-0.7162}^{+1.1991}}$\\
\\
\\
\hline
\end{tabular}
\end{center}
The rescaling factors for charged leptons are found as\\
${\eta_{\tau}= 1.017\pm .0007}$, ${\eta_{\mu}= 1.027\pm .0038}$, ${\eta_{e}= 1.046\pm .0099}$
for all cases.

For Case-III with the value of ${\alpha_{3}(M_{Z})=.118\pm .007}$,
the results estimated by Deshpande and Keith [12] 
using the same value of ${\alpha_{3}(M_{Z})=.118\pm .007}$, have small uncertainties as follows.\\

${\eta_{b}=1.53^{+0.07}_{-0.06}}$,
 ${\eta_{c}=2.09^{+0.27}_{-0.19}}$,
 ${\eta_{s,d}=2.36^{+0.53}_{-0.29}}$,
${\eta_{u}=2.38^{+0.52}_{-0.30}}$.\\ 

These low uncertainties may be resulted from an apparent mistake in their
estimation of uncertainties in\\
 ${\alpha_{3}^{-1}(\mu _{t})=9.30^{-0.047}_{+0.054}}$,\\
 instead, it should be corrected as \\ 
${\alpha_{3}^{-1}(\mu _{t})=9.30^{-0.47}_{+0.54}}$.\\
which will lead to larger uncertainties in $\eta_{f}$.

The present estimation of uncertainties in ${\eta_{f}}$ using quadrature method is very convincing and also it regulates 
the propagation of uncertainties while  running  the energy scale from high to low scales. These uncertainties can reliably 
be used in other low energy calculations e.g.,  neutrino masses and mixings in see-saw mechanism in low enery scale[13].\\

\section{Summary and conclusion:}
To summarise, we have outlined the procedure for estimating the uncertainties using the quadrature
method. In particular, we employ this
method to estimate the numerical uncertainties in QCD-QED rescaling factors ${\eta_{f}}$ acquired from the 
uncertainties in input values of ${\alpha_{3}(M_{Z})}$ and ${\alpha(M_{Z})}$ while running the 
energy scale from high to low. We have used the three-loop order in QCD and one-loop order in QED 
evolution equations to calculate the rescaling factors 
while their uncertainties estimated by the qudrature method are found to be new  in comparison 
to earlier estimates. 

\section*{Acknowledgements}
One of us (N.N.S) thanks Prof.M.K.Parida who  first highlited  this problem,  and 
the Abdus Salam International Centre for Theoretical Physics, Trieste, Italy, 
for kind hospitality during my visit at ICTP.

\pagebreak
\pagebreak

Table-1: Coefficients of ${\beta}$-functions in the 
RGEs for QED,the values of ${r_{f}^{QED}}$
in different energy ranges ${(\mu - \mu^{\prime})}$, ${\mu > \mu^{\prime}}$, the inverse of the gauge coupling
${\alpha^{-1}(\mu)}$ and inverse of the rescaling factors  ${R_{f}^{QED}(\mu,\mu^{\prime})}$ 
 for charged leptons ${(e,\mu ,\tau)}$ with input value
of ${\alpha^{-1}(M_{Z})=127.9\pm 0.1}$.
\begin{center}
\begin{tabular}{ccccc}\hline
Energy range                &  ${b_{0}^{QED}}$ & ${r_{f}^{QED}}$ &  ${\alpha ^{-1}(\mu)}$ & ${R^{f}_{QED}}$\\
${(\mu - \mu^{\prime})}$ &    &  ${(f=e,\mu ,\tau )}$  &     &  ${(f=e,\mu ,\tau )}$     \\
\hline
\\
${m_{t}-m_{b}}$     &       80/9  &  -0.3375 & ${126.98\pm 0.1}$  & ${0.9865 \pm 0.0002}$\\
\\
${m_{b}-m_{\tau}}$    &       79/9  &  -0.3553 & ${132.24\pm 0.1}$  & ${0.9968 \pm 0.0005}$\\
\\
${m_{\tau}-m_{c}}$    &       64/9  &  -0.4219 & ${133.40\pm 0.1}$  & ${0.9987\pm 0.0006}$\\
\\
${m_{c}-1GeV}$    &       16/3  &  -0.5625 & ${133.81\pm 0.1}$  & ${0.9990\pm 0.0008}$\\
\\
${1GeV-m_{s}}$    &       16/3  &  -0.5625 & ${133.99\pm 0.1}$  & ${0.9939 \pm 0.0007}$\\
\\
${m_{s}-m_{\mu}}$    &       44/9  &  -0.6136 & ${135.45\pm 0.1}$  & ${0.9981 \pm 0.0089}$\\
\\
${m_{\mu}-m_{d}}$    &       32/9  &  -0.8438 & ${135.87\pm 0.1}$  & ${0.9914 \pm 0.0011}$\\
\\
${m_{d}-m_{u}}$    &       29/9  &  -0.9643 & ${137.26\pm 0.1}$  & ${0.9979 \pm 0.0014}$\\
\\
${m_{u}-m_{e}}$    &       4/3  &  -2.2500 & ${137.55\pm 0.1}$  & ${0.9921 \pm 0.0032}$\\
\\
${m_{e}- 0}$   &  ...   & ...  &  ${138.04 \pm 0.1}$   &  ... \\
\\
\hline
\end{tabular}
\end{center}

\pagebreak
Table-2: Coefficients of ${\beta}$-functions in the 
RGEs for QED,the values of ${r_{f}^{QED}}$
in different energy ranges ${(\mu - \mu^{\prime})}$, ${\mu > \mu^{\prime}}$, and inverse of 
the rescaling factors  ${R_{f}^{QED}(\mu,\mu^{\prime})}$ for up- and down-quarks with 
renormalisation scale stopping at 1GeV.
  
\begin{center}
\begin{tabular}{cccccc}\hline
Energy scale   &  ${b_{0}^{QED}}$ & ${r_{f}^{QED}}$ &  ${r_{f}^{QED}}$ & ${R^{f}_{QED}}$ & ${R^{f}_{QED}}$\\
${(\mu - \mu^{\prime})}$ &    &  (f=u,c.t)  &  (f=s,d,b)   &  (f=u,ct)   & (f=s,d,b)  \\
\hline
\\
${m_{t}-m_{b}}$    &       80/9  &  -0.15 &  -0.0375  & ${0.9839 \pm 0.0002}$ & ${0.9985 \pm 0.0001}$\\
\\
${m_{b}-m_{\tau}}$    &       79/9  &  -0.1579 & -0.0395 &${0.9986\pm 0.0002}$  & ${0.9997 \pm 0.0001}$\\
\\
${m_{\tau}-m_{c}}$    &       64/9  &  -0.1875 & -0.0469 & ${0.9994\pm 0.0003}$  & ${0.9999\pm 0.0001}$\\
\\
${m_{c}-1GeV}$    &       16/3  &  -0.25  &  -0.0625 & ${0.9996\pm 0.0004}$  & ${0.9999\pm 0.0001}$\\
\\
\hline
\end{tabular}
\end{center}

Table-3: Coefficients of ${\beta}$ functions in RGEs, anomalous dimensions for QCD and values of the constants
A and B defined in Eq.(10) with ${\gamma_{0}=4}$ throughout the computation.

\begin{center}
\begin{tabular}{cccccccccc}\hline
Energy scale                &  ${\gamma_{1}}$ & ${\gamma_{2}}$ &  ${b_{0}}$ & ${b_{1}}$ &${b_{2}}$ & A & B & ${n_{f}}$ \\
${(\mu - \mu^{\prime})}$ &    &    &     &     &   &  &   & \\
\hline
\\
${m_{t}-m_{b}}$    & 506/9  & 474.89 & 23/3   & 116/3 & 9769/54 & 4.7020 & 24.0123 & 5\\
\\
${m_{b}-m_{\tau}}$    & 526/9  & 636.62 & 25/3 & 154/3 & 21943/54 & 4.0563 & 22.2280 & 4\\
\\
${m_{\tau}-m_{c}}$     & 526/9  & 636.62 & 25/3 & 154/3 & 21943/54 & 4.0563 & 22.2280 & 4\\
\\
${m_{c}-1GeV}$    &    546/9 & 794.90 & 9 & 64 & 34767/54 & 3.5802 & 21.9434 & 3\\
\\
\hline
\end{tabular}
\end{center}

\pagebreak
Table-4: The values of QCD dimensional parameters ${\Lambda^{f}_{MS}}$ with uncertainties 
estimated for three different values
of ${\alpha_{3}(\mu)}$ for cases I, II, III.
\begin{center}
\begin{tabular}{cccc}\hline
case &  ${\Lambda^{(5)}_{MS}}$GeV &  ${\Lambda^{(4)}_{MS}}$GeV &  ${\Lambda^{(3)}_{MS}}$GeV \\
\hline
\\
I   & ${0.1995_{-0.0217}^{+0.0243}}$ & ${0.2877^{+0.0313}_{-0.0283}}$ & ${0.3375^{+0.0325}_{-0.0295}}$ \\
\\
II & ${0.2329_{-0.0334}^{+0.0364}}$ &  ${0.3305_{-0.0429}^{+0.0459}}$ & ${0.3838_{-0.0448}^{+0.0472}}$\\
\\
III & ${0.2087_{-0.0716}^{+0.0934}}$ & ${ 0.2995_{-0.0942}^{+0.1177}}$ & ${0.3498_{-0.1001}^{+0.1191}}$\\
\\
\hline
\end{tabular}
\end{center}

Table-5: The values of ${\alpha_{3}(\mu)}$ and ${R_{QCD}^{f}(\mu - \mu^{\prime})}$ with uncertainties estimated 
for quarks using three-loop
order in QCD evolution equations for three different input experimental values of ${\alpha_{3}(M_{Z})}$.
Here QM stands for the quadrature method employed here.

\begin{center}
\begin{tabular}{cccccc}\hline
 case /parameter & ${m_{t}-m_{b}}$ & ${m_{b}-m_{\tau}}$  &${m_{\tau}-m_{c}}$  & ${m_{c}-1GeV}$ & ${1GeV-0}$     
\\
\hline
\\
I /${\alpha_{3}(\mu)}$ & ${0.1068_{-0.0016}^{+0.0017}}$  &${0.2225_{-0.0073}^{+0.0079}}$ & 
${0.3149_{-0.0152}^{+0.0170}}$ & ${0.3967_{-0.0254}^{+0.0303}}$ & ${0.4819_{-0.0400}^{+0.0507}}$ \\
\\
(QM)/ ${R_{QCD}^{f}}$ & ${0.6512_{-0.0142}^{+0.0153}}$ & ${0.8185_{-0.0279}^{+0.0310}}$ & 
${0.8682_{-0.0430}^{+0.0503}}$ & ${0.8900_{-0.0564}^{+0.0700}}$ & 
\\
\hline 
\\
II/${\alpha_{3}(\mu)}$ & ${0.1091_{-0.0023}^{+0.0022}}$  &${0.2334_{-0.0107}^{+0.0112}}$ & 
${0.3382_{-0.0234}^{+0.0257}}$ & ${0.4233_{-0.0389}^{+0.0466}}$ & ${0.5568_{-0.0728}^{+0.1001}}$ \\

\\ 
(QM)/${R_{QCD}^{f}}$ & ${0.6400_{-0.0196}^{+0.0206}}$ & ${0.8055_{-0.0396}^{+0.0430}}$ & 
${0.8696_{-0.0629}^{+0.0732}}$ & ${0.8439_{-0.0851}^{+0.1129}}$ & \\
\\
\hline
\\ 
III/${\alpha_{3}(\mu)}$ & ${0.1075_{-0.0059}^{+0.0057}}$  &${0.2257_{-0.0253}^{+0.0285}}$ & 
${0.3213_{-0.0508}^{+0.0663}}$ & ${0.4079_{-0.0828}^{+0.1334}}$ & ${0.5001_{-0.1254}^{+0.2652}}$ \\
\\
(QM)/${R_{QCD}^{f}}$ & ${0.6479_{-0.0489}^{+0.0538}}$ & ${0.8152_{-0.0927}^{+0.1161}}$ & 
${0.8633_{-0.1348}^{+0.2030}}$ & ${0.8753_{-0.1722}^{+0.3351}}$ & 
\\
\hline

\end{tabular}
\end{center}

\end{document}